\documentclass[11pt,onecolumn,english,preprintnumbers,amsmath,amssymb,floatfix,nofootinbib]{revtex4}
\usepackage{tikz,xcolor}
\usepackage[colorlinks = true,  
linkcolor = blue,
urlcolor  = blue,
citecolor = blue,
anchorcolor = blue]{hyperref}

\definecolor{lime}{HTML}{A6CE39}
\DeclareRobustCommand{\orcidicon}{%
	\begin{tikzpicture}
		\draw[lime, fill=lime] (0,0) 
		circle [radius=0.16] 
		node[white] {{\fontfamily{qag}\selectfont \tiny ID}};
		\draw[white, fill=white] (-0.0625,0.095) 
		circle [radius=0.007];
	\end{tikzpicture}
	\hspace{-2mm}
}

\foreach \x in {A, ..., Z}{%
	\expandafter\xdef\csname orcid\x\endcsname{\noexpand\href{https://orcid.org/\csname orcidauthor\x\endcsname}{\noexpand\orcidicon}}
}

\usepackage[T1]{fontenc}
\usepackage[latin9]{inputenc}
\usepackage{color}
\usepackage{array}
\usepackage{amstext}
\usepackage{graphicx}
\usepackage{esint}
\usepackage{rotating}
\usepackage{appendix}
\usepackage{float}
\usepackage{subcaption}
\usepackage{amsmath}

\usepackage[font=small,labelfont=bf]{caption}
\usepackage{xcolor}
\usepackage{ulem}
\usepackage{amssymb}

\makeatletter

\@ifundefined{textcolor}{}
{%
	\definecolor{BLACK}{gray}{0}
	\definecolor{WHITE}{gray}{1}
	\definecolor{RED}{rgb}{1,0,0}
	\definecolor{GREEN}{rgb}{0,1,0}
	\definecolor{BLUE}{rgb}{0,0,1}
	\definecolor{CYAN}{cmyk}{1,0,0,0}
	\definecolor{MAGENTA}{cmyk}{0,1,0,0}
	\definecolor{YELLOW}{cmyk}{0,0,1,0}
}

\@ifundefined{definecolor}
{\usepackage{color}}{}
\@ifundefined{definecolor}
{\usepackage{color}}{}
\makeatother
\usepackage{babel}

\newcommand{\beq}{\begin{equation}}
	\newcommand{\eeq}{\end{equation}}

\def\Re{{\cal R \mskip-4mu \lower.1ex \hbox{\it e}\,}}
\def\Im{{\cal I \mskip-5mu \lower.1ex \hbox{\it m}\,}}

\def\tev{\,{\ifmmode\mathrm {TeV}\else TeV\fi}}
\def\gev{\,{\ifmmode\mathrm {GeV}\else GeV\fi}}
\def\mev{\,{\ifmmode\mathrm {MeV}\else MeV\fi}}
\def\to{\rightarrow}

\begin{document}
		
	\title{ Probing anomalous $hZ\gamma$ couplings in single Higgs production at future muon colliders }

	\author{Shirin Chenarani$^{1,2}$\orcidA{}}
	\email{shirin.chenarani@cern.ch}

	\author{Sara Khatibi$^{3}$\orcidB{}}
	\email{sara.khatibi@ut.ac.ir}
	
	\affiliation {
		$^{(1)}$ Department of Physics, University of Science and Technology of Mazandaran, 48518-78195, Behshahr, Iran  \\
		$^{(2)}$ School of Particles and Accelerators, Institute for Research in Fundamental Sciences (IPM), 19395-5531, Tehran, Iran\\
		$^{(3)}$Department of Physics, University of Tehran, North Karegar Ave., Tehran 14395-547, Iran 
	}

		
\begin{abstract}
\label{abstract}
 Future multi-TeV muon colliders offer an ideal environment to probe physics beyond the Standard Model (SM)
 through high-precision measurements. In this work, we investigate the sensitivity to anomalous $hZ\gamma$ 
 couplings within the dimension-six Standard Model Effective Field Theory (SMEFT) framework 
 at $\sqrt{s} = 3\text{ TeV}$ and $10\text{ TeV}$ muon colliders. Focusing on single Higgs boson 
 production---particularly through vector boson fusion---followed by the rare $h \to Z\gamma$ decay channel, 
 we perform a comprehensive simulation using \textsc{MadGraph5\_aMC@NLO}, \textsc{Pythia}, and \textsc{Delphes} 
 for fast detector response. To maximize signal sensitivity, we contrast a traditional cut-and-count 
 methodology with a multi-bin Boosted Decision Tree (BDT) shape analysis, evaluated within a 
 multi-bin $\chi^2$ statistical framework. Our most stringent projections are achieved using 
 the multivariate BDT analysis at $10\text{ TeV}$. Assuming an integrated luminosity 
 of $10\text{ ab}^{-1}$ and a $5\%$ systematic uncertainty, we derive expected $95\%$ confidence level 
 intervals of $c_{HW} \in [-0.005, +0.003]$ and $c_{HB} \in [-0.003, +0.005]$. Notably, the bound 
 on the $c_{\gamma}$ coefficient is slightly tighter due to its larger cross-section enhancement, 
 reaching $c_{\gamma} \in [-0.002, 0.001]$. These results underscore the unique capability of 
 high-energy muon colliders to constrain anomalous Higgs-gauge couplings.
\end{abstract}  


	\maketitle


	\newpage
	\section{Introduction} \label{intro}
   	
  Since the ATLAS and CMS collaborations discovered the Higgs boson~\cite{ATLAS:2012yve,CMS:2012qbp}, 
its measured properties have consistently aligned with Standard Model (SM) predictions~\cite{ATLAS:2022vkf,CMS:2022dwd}. 
Nonetheless, rare decay channels like $h \to Z\gamma$ have not yet been formally observed, offering a highly 
sensitive probe to search for signs of physics beyond the SM (BSM). In the SM, the $h \to Z\gamma$ transition 
is loop-induced, resulting in a heavily suppressed branching fraction predicted to be 
approximately $(1.5 \pm 0.1) \times 10^{-3}$ for a Higgs mass of 
$125~\mathrm{GeV}$~\cite{Djouadi:1997yw,LHCHiggsCrossSectionWorkingGroup:2016ypw}. 
Consequently, this channel is highly sensitive to BSM physics, as new heavy virtual particles running in the 
loop can significantly enhance the decay rate compared to the 
SM expectation~\cite{Azatov:2013ura,Cao:2018cms,Low:2011gn,Low:2012rj,Carena:2012xa,Chiang:2012qz,Chen:2013vi}.

A recent combination of Run 2 datasets from the ATLAS and CMS collaborations provided the first evidence 
for this decay, reaching an observed significance of $3.4 \sigma$~\cite{ATLAS:2023yqk}. Notably, 
the measured signal strength was found to be $\mu = 2.2 \pm 0.7$, pointing to a slight elevation over the 
SM prediction. While still statistically limited, this mild tension, combined with the lack 
of a formal $5\sigma$ observation, strongly motivates the systematic search for anomalous $hZ\gamma$ couplings. 
Future high-energy facilities, particularly a multi-TeV muon collider, offer an exceptionally clean environment 
to scrutinize these electroweak interactions with unprecedented precision.

Anomalous $hZ\gamma$ couplings have been studied extensively across 
various future collider scenarios. Earlier work largely focused on future $e^+e^-$ factories~\cite{Cao:2015iua} 
and the High-Luminosity LHC~\cite{No:2016ezr, Goertz:2019uek}, utilizing channels like associated production 
and boosted Higgs topologies. Comprehensive global EFT analyses have further demonstrated the capacity of these future machines 
to systematically probe the Higgs sector with high precision~\cite{Durieux:2017rsg}. More recently, however, attention has shifted 
toward the multi-TeV frontier. High-energy muon colliders offer distinct advantages here, such as the ability to break 
parameter degeneracies, measure the total Higgs width, and determine the exact sign of gauge-Higgs 
couplings~\cite{Forslund:2023reu,deLima:2024ybb,Spor:2025ezn}. 

A multi-TeV muon collider offers a unique combination of high energy and experimental precision. 
Since muons are fundamental leptons, the full center-of-mass energy is available in the hard collision, 
allowing a $10~\mathrm{TeV}$ machine to match the physics reach of a $100~\mathrm{TeV}$ proton collider. 
This energy advantage is paired with a clean environment naturally free of heavy QCD backgrounds, 
enabling high-precision measurements~\cite{AlAli:2021let,Aime:2022flm}.
Furthermore, at multi-TeV scales, the rapid rise of vector boson fusion (VBF) cross sections effectively 
turns the collider into a vector boson factory, providing an ideal platform for probing rare Higgs processes~\cite{Costantini:2020stv}.

This paper explores the potential of a future multi-TeV muon collider to probe anomalous $hZ\gamma$ interactions 
through the Standard Model Effective Field Theory (SMEFT) framework. Specifically, we analyze 
the vector boson fusion channel $\mu^+\mu^- \to h\nu\bar{\nu}$ at center-of-mass energies of $\sqrt{s} = 3$ and $10~\mathrm{TeV}$, 
corresponding to benchmark integrated luminosities of $1~\mathrm{ab}^{-1}$ and $10~\mathrm{ab}^{-1}$, respectively. 
Focusing on the $h \to Z\gamma \to \ell^+\ell^-\gamma$ decay mode, we establish a baseline cut-and-count analysis 
and complement it with an optimized multi-bin Boosted Decision Tree (BDT) framework to project $95\%$ CL limits 
on the SILH basis dimension-six Wilson coefficients ($c_{HW}, c_{HB}, c_\gamma$). Finally, these bounds are mapped 
directly onto the physical effective coupling $g_{hZ\gamma}^{\text{eff}}$ to yield a basis-independent measure of collider sensitivity.

This paper is structured as follows. In Section~2, we establish the theoretical framework describing the anomalous 
$hZ\gamma$ couplings. Section~3 details our simulation, including Monte Carlo event generation, detector simulation, 
and the criteria for separating the signal from SM backgrounds. We present our sensitivity projections and results 
in Section~4, and conclude with a summary of our findings in Section~5.

\section{Theoretical Framework}\label{sec:model}

Given the remarkable agreement between current experimental measurements and SM predictions, 
the scale of any new physics is likely situated well above our current kinematic reach. In this regime, 
the SMEFT offers an elegant, model-independent framework to search 
for these heavy degrees of freedom. By systematically integrating out the heavy fields, SMEFT captures 
their indirect effects through higher-dimensional operators constructed entirely from SM fields while respecting 
Lorentz and $SU(3)_C \times SU(2)_L \times U(1)_Y$ gauge symmetries. Assuming baryon and lepton number conservation, 
the lowest-dimensional modifications arise at dimension-six, suppressed by two powers of the new physics scale, $\Lambda$. 

Following the SILH basis conventions~\cite{Contino:2013kra,Alloul:2013naa}, the effective Lagrangian 
expanded up to dimension-six operators is expressed as:
\begin{equation}
\mathcal{L}_{\mathrm{SMEFT}} = \mathcal{L}_{\mathrm{SM}} + \sum_i c_i \mathcal{O}_i, 
\label{eq:smeft_lagrangian}
\end{equation}
where $\mathcal{L}_{\mathrm{SM}}$ represents the renormalizable SM Lagrangian, $\mathcal{O}_i$ denote the dimension-six 
operators, and $c_i$ are the corresponding dimensionless Wilson coefficients that parameterize the strength of these 
effective interactions relative to the electroweak scale.

The specific CP-conserving dimension-six operators that drive modifications to the $h \to Z\gamma$ decay channel 
are given by:
\begin{align}
\mathcal{O}_{HW} &= \frac{i g}{m_W^2} (D^\mu H)^\dagger \sigma^I (D^\nu H) W^I_{\mu\nu} , \nonumber \\
\mathcal{O}_{HB} &= \frac{i g'}{m_W^2} (D^\mu H)^\dagger (D^\nu H) B_{\mu\nu} , \nonumber \\
\mathcal{O}_\gamma &= \frac{g'^2}{m_W^2} |H|^2 B_{\mu\nu} B^{\mu\nu} ,
\label{eq:operators}
\end{align}
where $H$ denotes the SM Higgs doublet, $W^I_{\mu\nu}$ ($I=1,2,3$) and $B_{\mu\nu}$ represent the $SU(2)_L$ and $U(1)_Y$ 
electroweak field strength tensors, $g$ and $g'$ are the respective gauge couplings, and $\sigma^I$ are the Pauli matrices. 

After electroweak symmetry breaking (EWSB), the dimension-six operators in Eq.~\eqref{eq:operators} 
generate an effective $hZ\gamma$ interaction vertex at tree level. 
The corresponding Lagrangian contributing to the $h \to Z\gamma$ decay is given by:
\begin{equation}
\mathcal{L}_{hZ\gamma}^{\text{eff}} = \frac{g_{hZ\gamma}^{\text{eff}}}{v} h Z_{\mu\nu} A^{\mu\nu},
\label{eq:L_hZgamma}
\end{equation}
where $Z_{\mu\nu}$ and $A_{\mu\nu}$ are the field strength tensors of the $Z$ boson and photon, 
and $v \approx 246~\mathrm{GeV}$ is the Higgs vacuum expectation value. The effective 
coupling $g_{hZ\gamma}^{\text{eff}}$ parameterizes the net shift induced by the dimension-six 
operators and is expressed in terms of the Wilson coefficients as:
\begin{equation}
g_{hZ\gamma}^{\text{eff}} = -\tan\theta_W (c_{HW} - c_{HB}) + 8 \sin^2\theta_W c_\gamma,
\label{eq:g_hzgamma}
\end{equation}
where $\theta_W$ represents the weak mixing angle.

In the following section, the effective Lagrangian introduced above is employed to investigate the $h \to Z\gamma$ decay mode, 
focusing on Higgs production in association with missing transverse energy (via $\mu^+\mu^- \to \nu\bar{\nu}h$) at a 
future high-energy muon collider. Crucially, these SMEFT operators are considered exclusively in the Higgs decay vertex, 
while the vector boson fusion production process is treated as SM-like.  Furthermore, we examine these three operators 
individually in our analysis---activating one Wilson coefficient at a time while setting the others to zero---to extract 
individual projected $95\%$ CL bounds on $c_{HW}$, $c_{HB}$, and $c_{\gamma}$.
		
\section{Analysis strategy}
\label{sec:analysis_strategy}

In this section, we present our strategy for probing the $h \to Z\gamma$ decay mode in $\mu^+\mu^- \to \nu\bar{\nu}h$ 
production at a future high-energy muon collider. We first define the signal and dominant SM background channels, 
followed by a detailed description of the Monte Carlo event generation and detector simulation framework. 
The analysis is performed at two benchmark center-of-mass energies: $\sqrt{s} = 3~\mathrm{TeV}$ and $10~\mathrm{TeV}$, 
assuming integrated luminosities of $1~\mathrm{ab}^{-1}$ and $10~\mathrm{ab}^{-1}$, respectively.
Throughout this analysis, both initial-state muon and antimuon beams are taken to be unpolarized.

\subsection{Event Generation and Detector Simulation}
\label{sec:simulation}

\begin{figure}[htbp]
\centering
\includegraphics[width=0.23\textwidth]{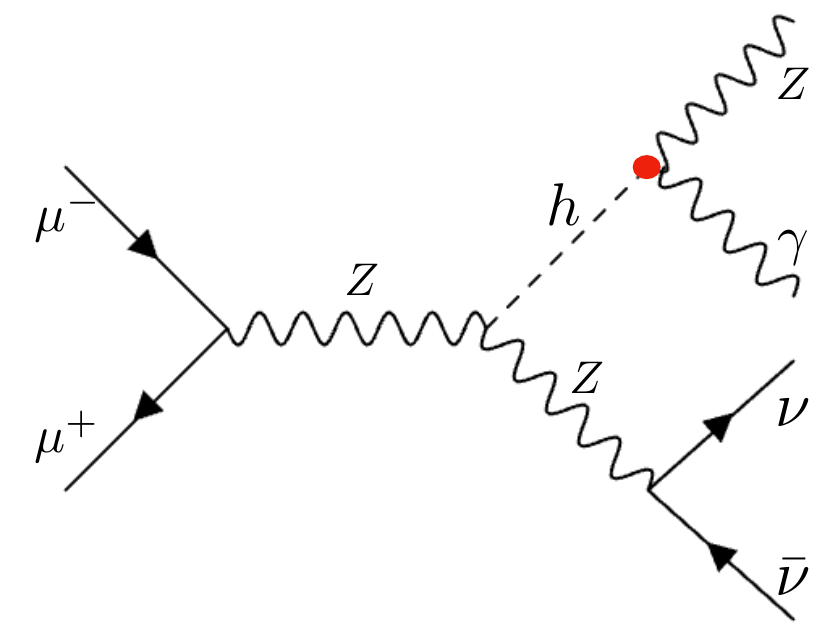}\hfill
\includegraphics[width=0.20\textwidth]{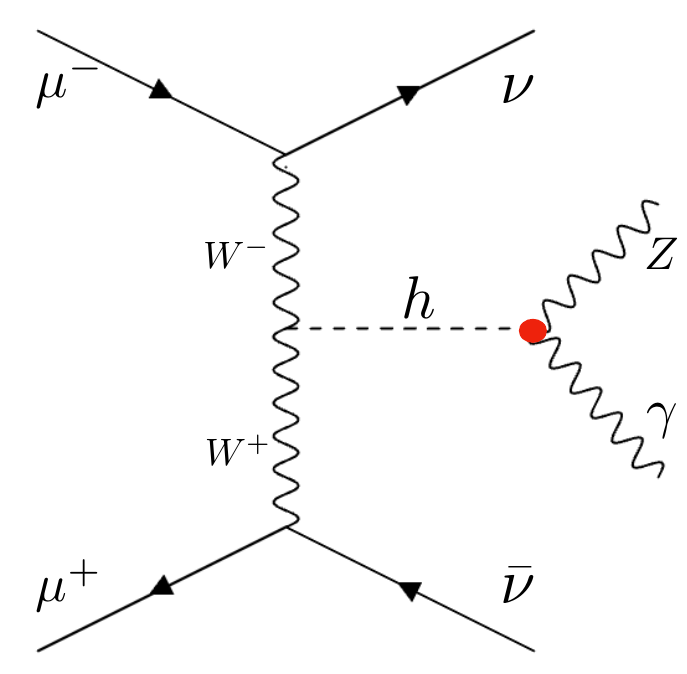}\hfill
\includegraphics[width=0.20\textwidth]{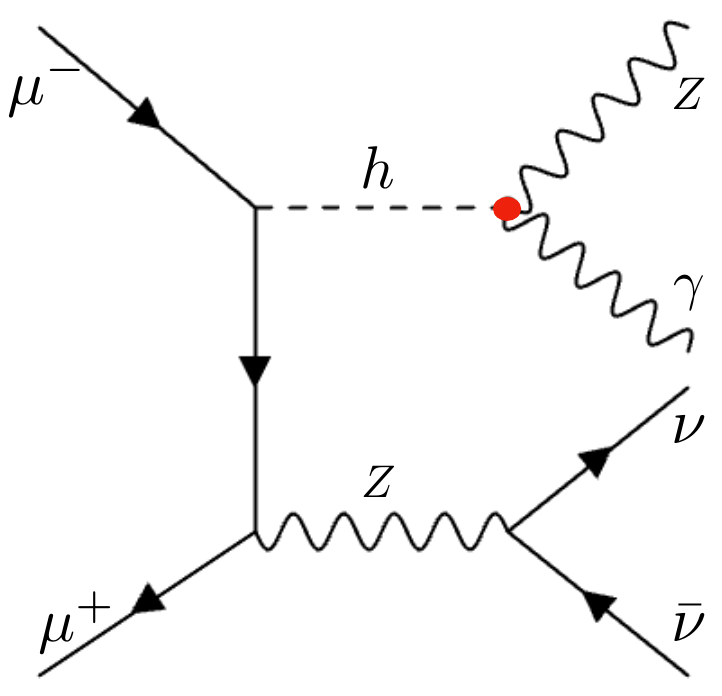}\hfill
\includegraphics[width=0.17\textwidth]{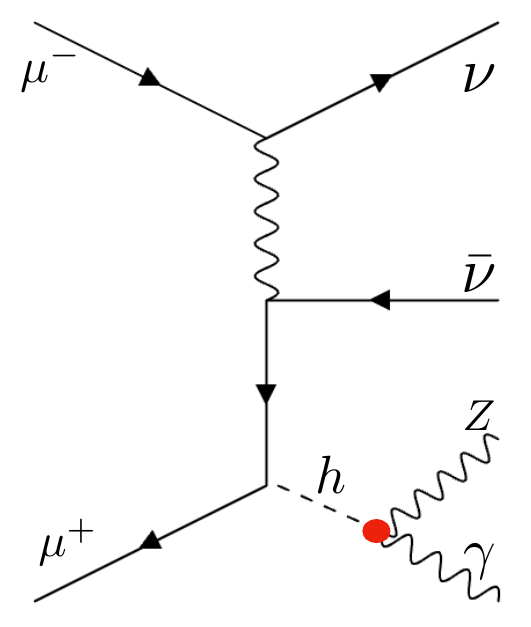}
\caption{Representative Feynman diagrams for the signal process $\mu^+\mu^- \to \nu\bar{\nu}h (\to Z\gamma)$ at a high-energy muon collider.}
\label{fig:feynman}
\end{figure}

We define the signal process as vector boson fusion (VBF) Higgs production, $\mu^+ \mu^- \to \nu\bar{\nu}h$, 
followed by the sequential decays $h \to Z\gamma$ and $Z \to \ell^+ \ell^-$, where $\ell \in \{e, \mu\}$ and 
the neutrino sum includes all three SM flavors ($\nu_e, \nu_\mu, \nu_\tau$). The primary experimental signature 
thus comprises exactly two opposite-sign same-flavor (OSSF) charged leptons, one isolated high-$p_T$ photon, 
and missing transverse energy ($E_T^{\text{miss}}$) carried away by the escaping neutrinos. 
Representative Feynman diagrams for the signal process are illustrated in Fig.~\ref{fig:feynman}.

To evaluate sensitivity, we simulate several SM background processes yielding identical or kinematically 
similar $\ell^+\ell^-\gamma + E_T^{\text{miss}}$ final states. The dominant irreducible contribution arises 
from $\mu^+\mu^- \to \nu\bar{\nu}Z\gamma$ with leptonic $Z$ decay. Significant reducible background sources 
are also included: resonant triboson processes such as $ZZ\gamma$ production (with one $Z \to \ell^+\ell^-$ 
and one $Z \to \nu\bar{\nu}$) and leptonic $W^+W^-\gamma$ production ($W \to \ell \nu$), as well as the 
electroweak ditau channel $\mu^+\mu^- \to \tau^+\tau^-\gamma$ in which both taus decay leptonically 
($\tau \to \ell \nu_\ell \nu_\tau$), inherently mimicking the missing energy signal signature.

Signal events containing the dimension-six operators in Eq.~\eqref{eq:operators} are generated using the \texttt{HEL\_UFO} 
model framework~\cite{Alloul:2013naa,Degrande:2011ua}, which we modified to explicitly include the loop-induced S
M $h \to Z\gamma$ decay vertex alongside the higher-dimensional operator contributions. The modified UFO model 
is imported into \textsc{MadGraph5\_aMC@NLO}~\cite{Alwall:2011uj} to generate signal events. In our analysis, 
we examine these three operators individually by varying one Wilson coefficient at a time while setting the 
others to zero. SM background samples are likewise generated at leading order (LO) in \textsc{MadGraph5\_aMC@NLO}. 
At the generator level, loose preselection cuts are applied to all visible final-state particles ($p_T \ge 10\text{ GeV}$ 
and $|\eta| < 2.5$) to ensure initial acceptance. Hard-scattering parton-level events are subsequently showered 
and hadronized using \textsc{Pythia}~8.3~\cite{Bierlich:2022pfr}. Detector response and object reconstruction 
are simulated through \textsc{Delphes}~3~\cite{deFavereau:2013fsa}, utilizing the dedicated muon collider 
detector performance card~\cite{Accettura:2023,MuonColliderDelphesCard}.

To account for the experimental impact of the beam-induced background (BIB) from decaying muons and the 
physical constraints of forward tungsten shielding nozzles, all reconstructed objects are restricted to 
the fiducial volume $\vert{}\eta\vert{} \le 2.4$ with a transverse momentum threshold of $p_T \ge 20\text{ GeV}$. 
Under these acceptance criteria, residual BIB contributions can be safely neglected. Candidate events are then 
required to contain at least two OSSF leptons and at least one isolated photon satisfying an angular separation 
of $\Delta R(\ell, \gamma) > 0.4$. The $Z$-boson candidate is reconstructed from the OSSF lepton pair within 
the invariant mass window $76\text{ GeV} < m_{\ell\ell} < 106\text{ GeV}$, and the candidate Higgs boson is 
subsequently formed by combining the $Z$ candidate with the leading photon.

\subsection{Signal-Background Discrimination}
\label{sec:bdt}

\begin{figure*}[htbp]
	\centering
	\includegraphics[width=0.45\textwidth]{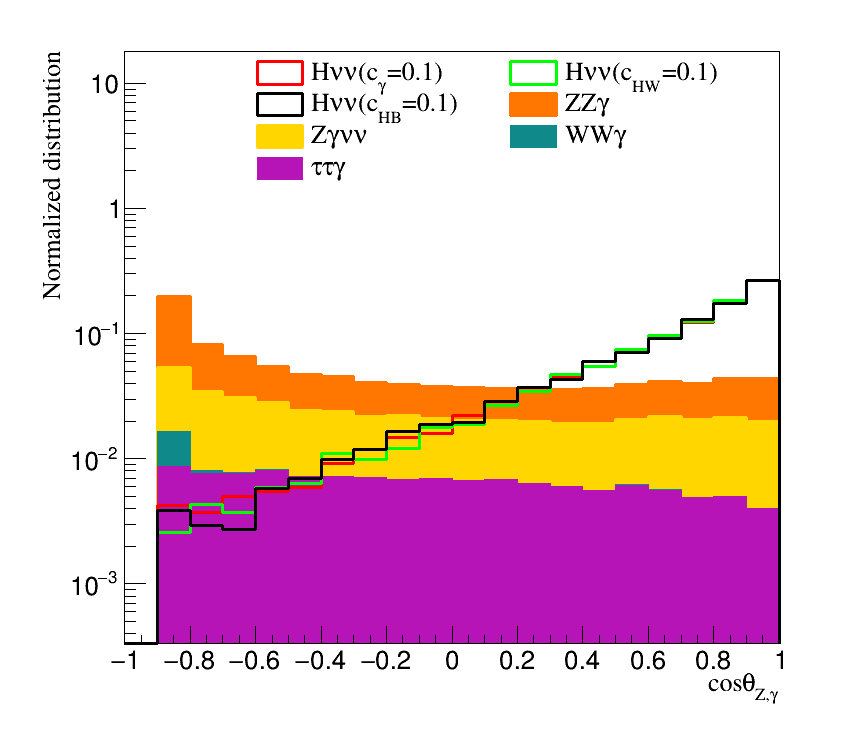}
	\includegraphics[width=0.45\textwidth]{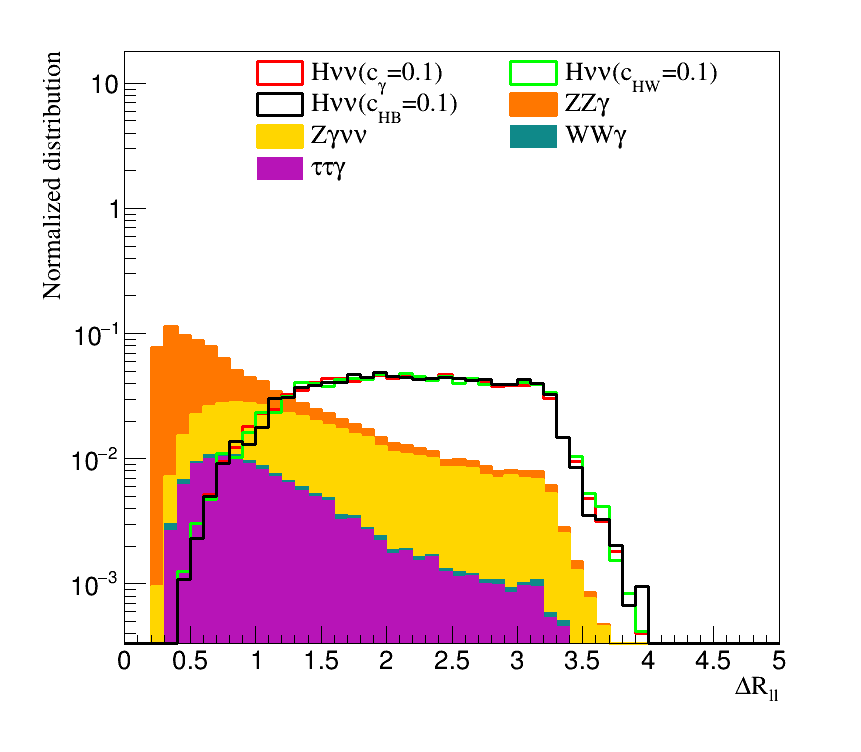}  \\   
	\includegraphics[width=0.45\textwidth]{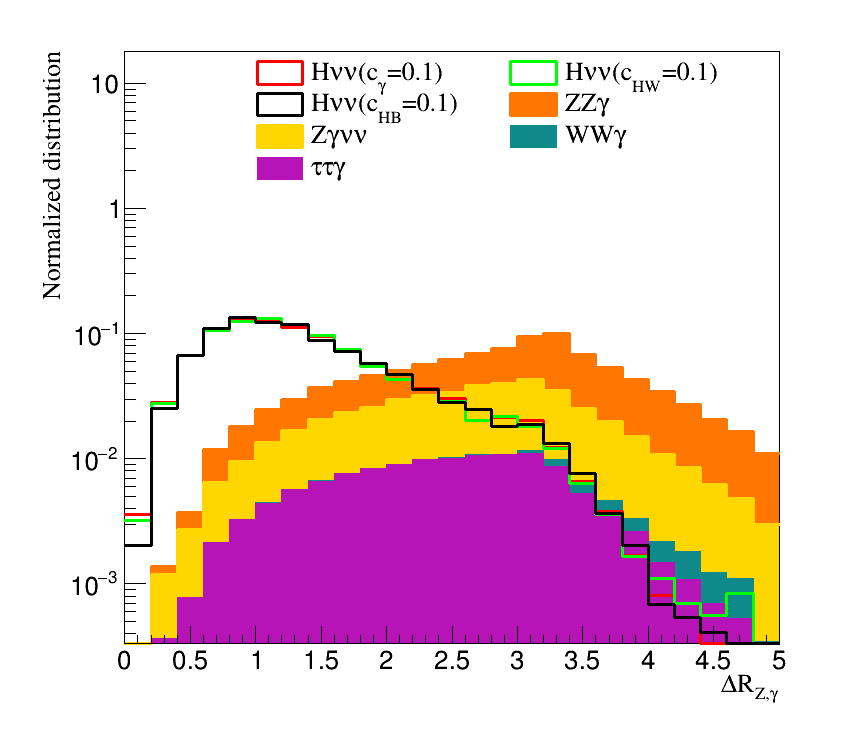}
	\includegraphics[width=0.45\textwidth]{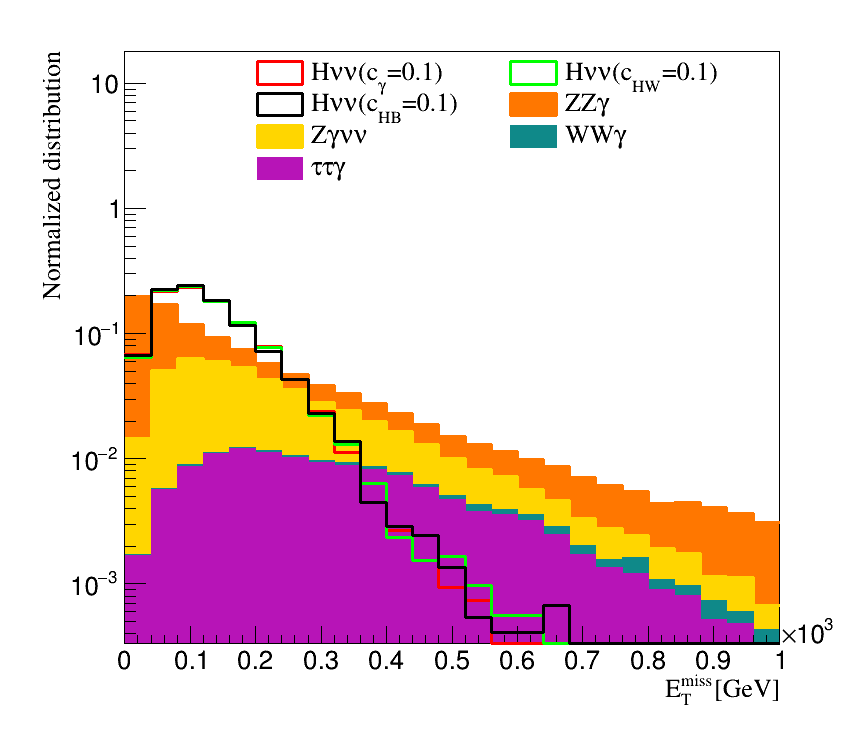}\\
	\includegraphics[width=0.45\textwidth]{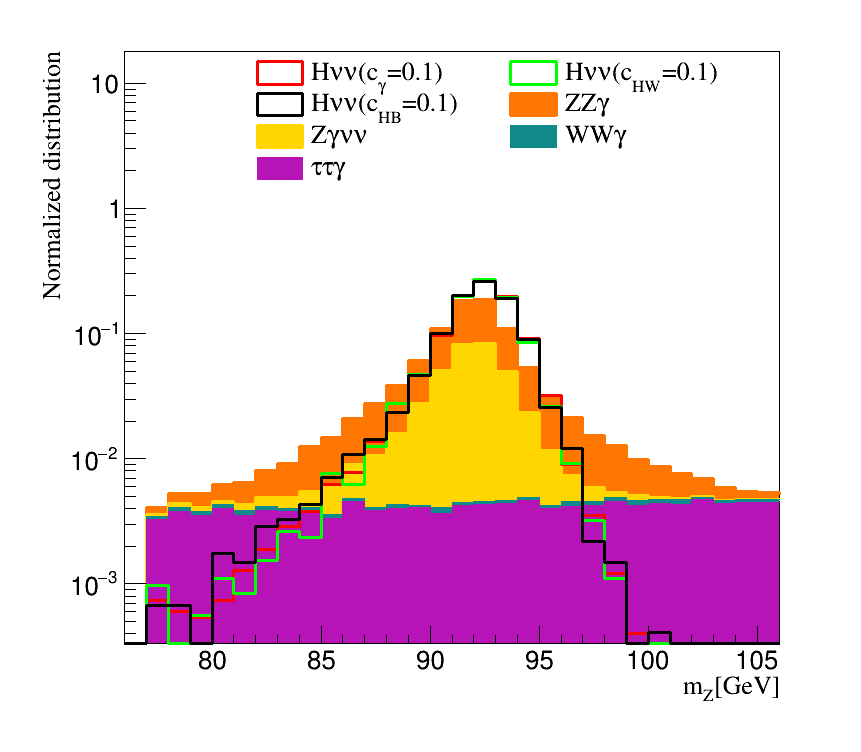}     
	\includegraphics[width=0.45\textwidth]{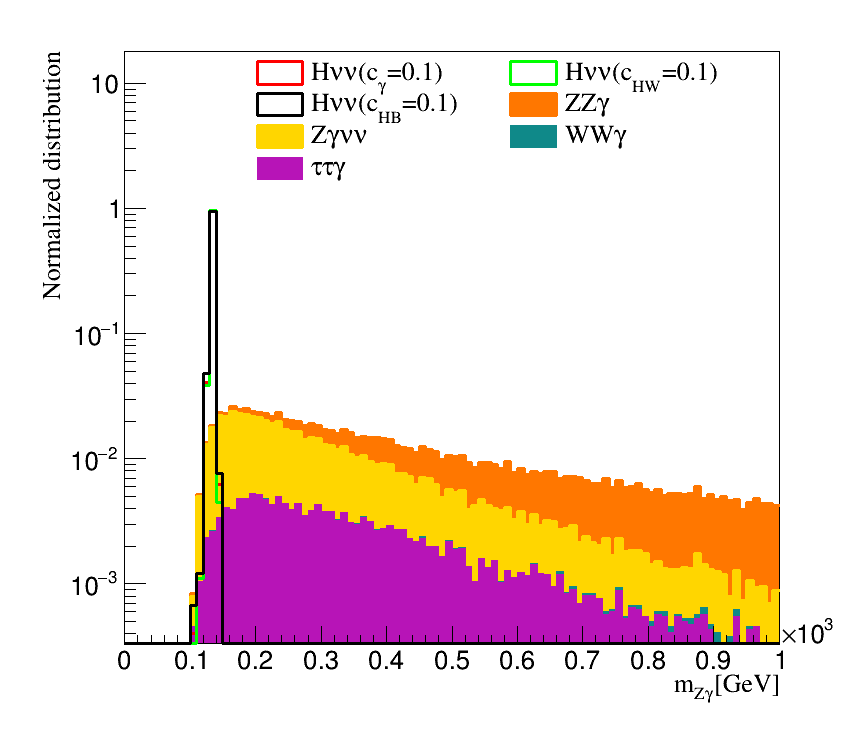}
	\caption{Normalized kinematic distributions of input observables utilized for 
	multivariate signal-background discrimination at $\sqrt{s} = 3\text{ TeV}$.}
	\label{fig:Variables-Distributions3}
\end{figure*}

\begin{figure*}[htbp]
	\centering
	\includegraphics[width=0.45\textwidth]{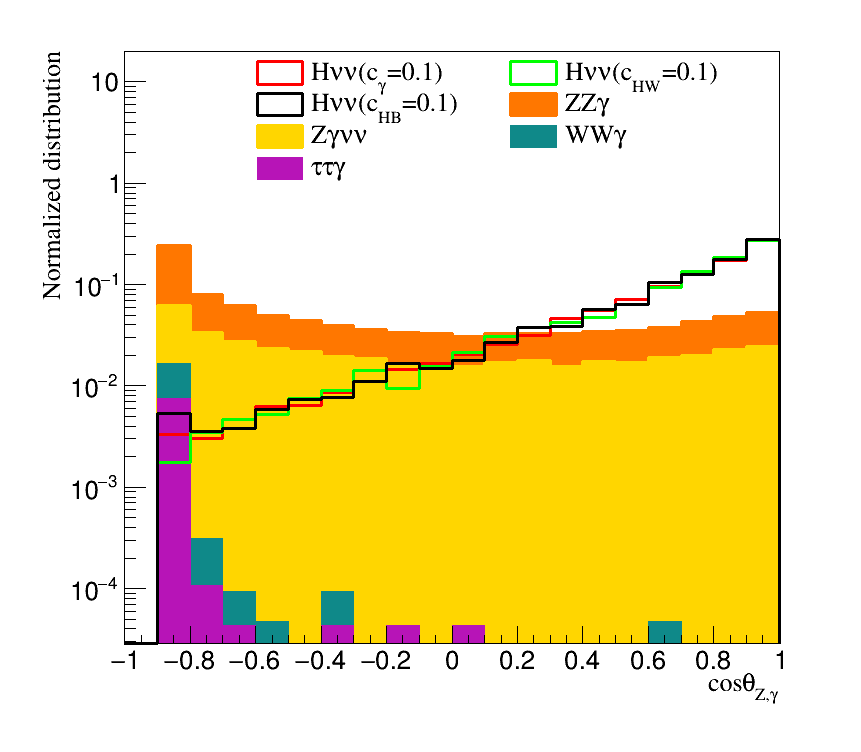}
	\includegraphics[width=0.45\textwidth]{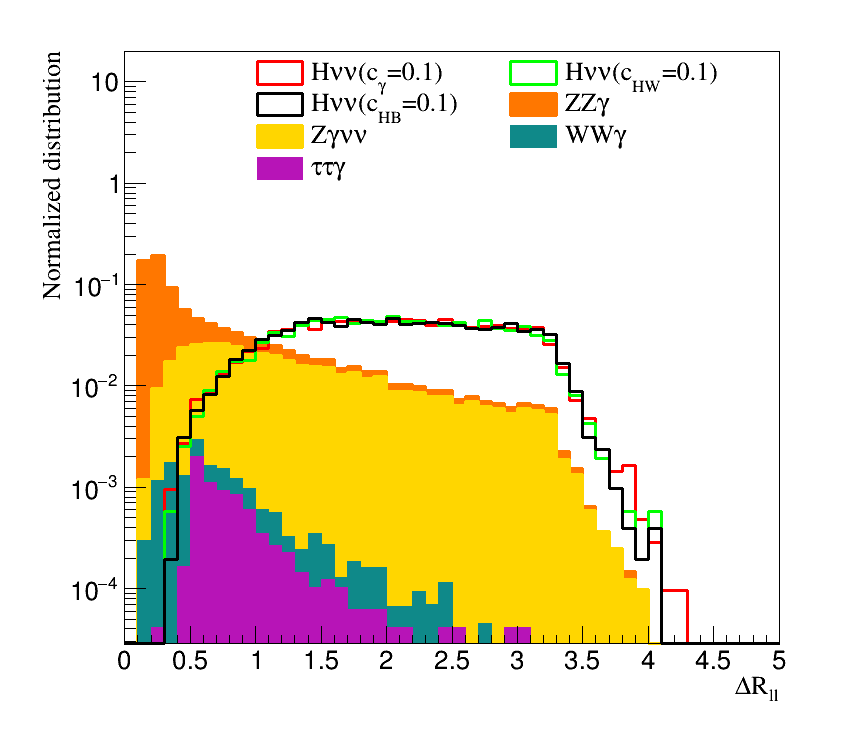}   \\  
	\includegraphics[width=0.45\textwidth]{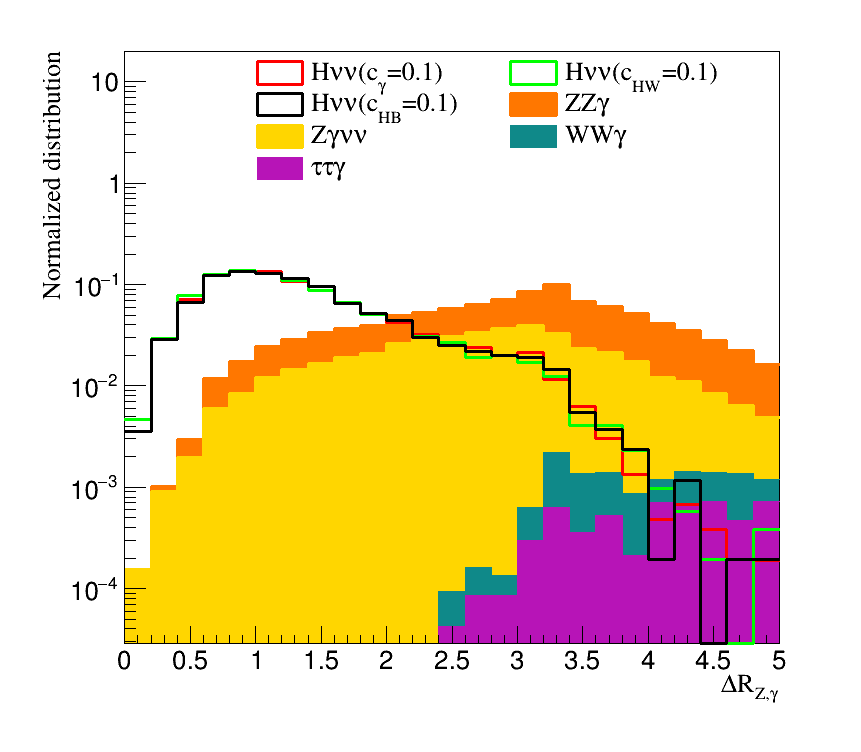}
	\includegraphics[width=0.45\textwidth]{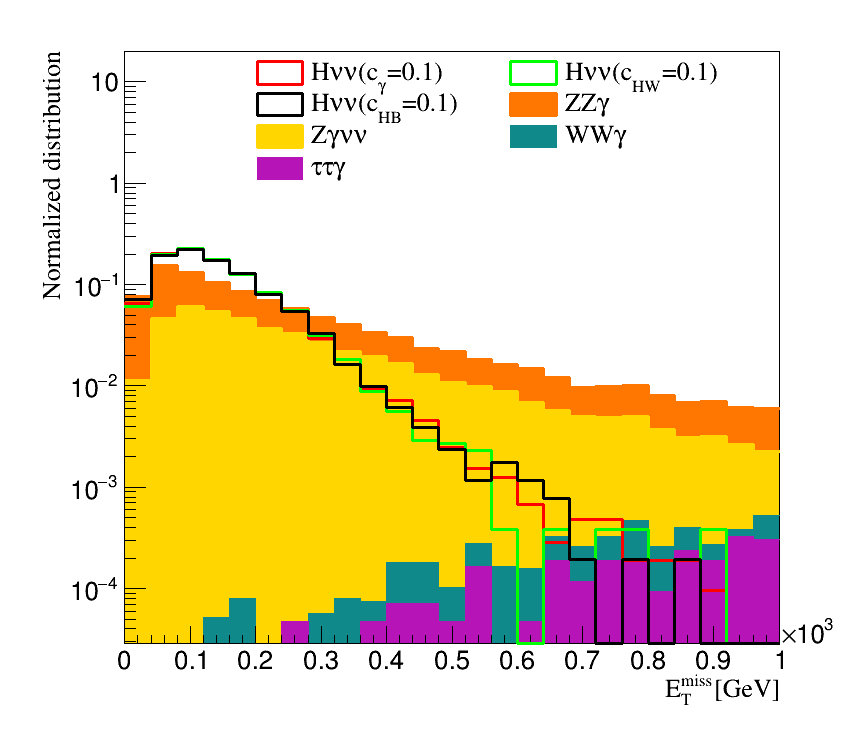}\\
	\includegraphics[width=0.45\textwidth]{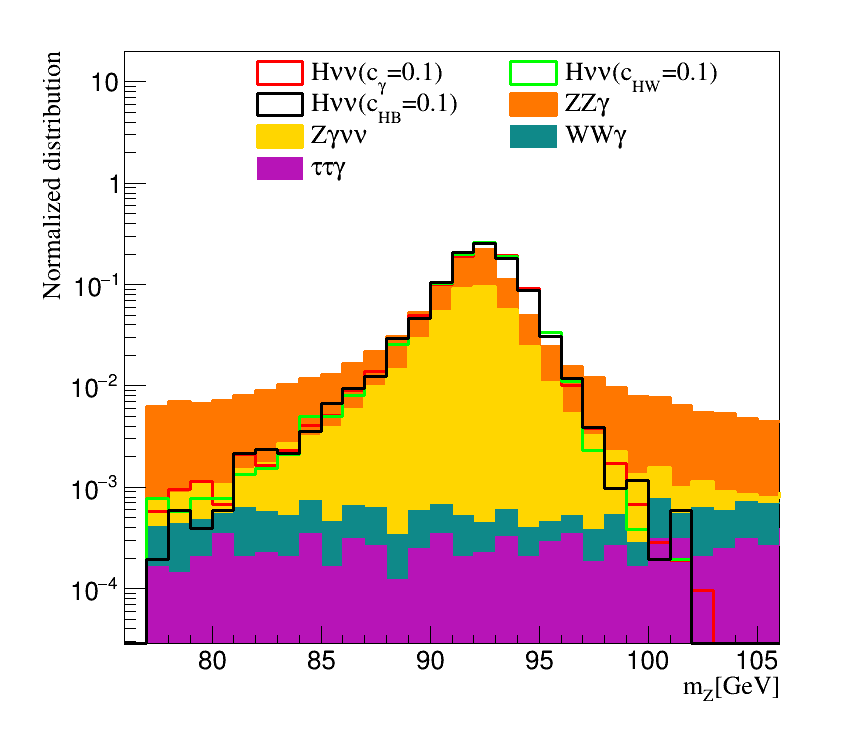}     
	\includegraphics[width=0.45\textwidth]{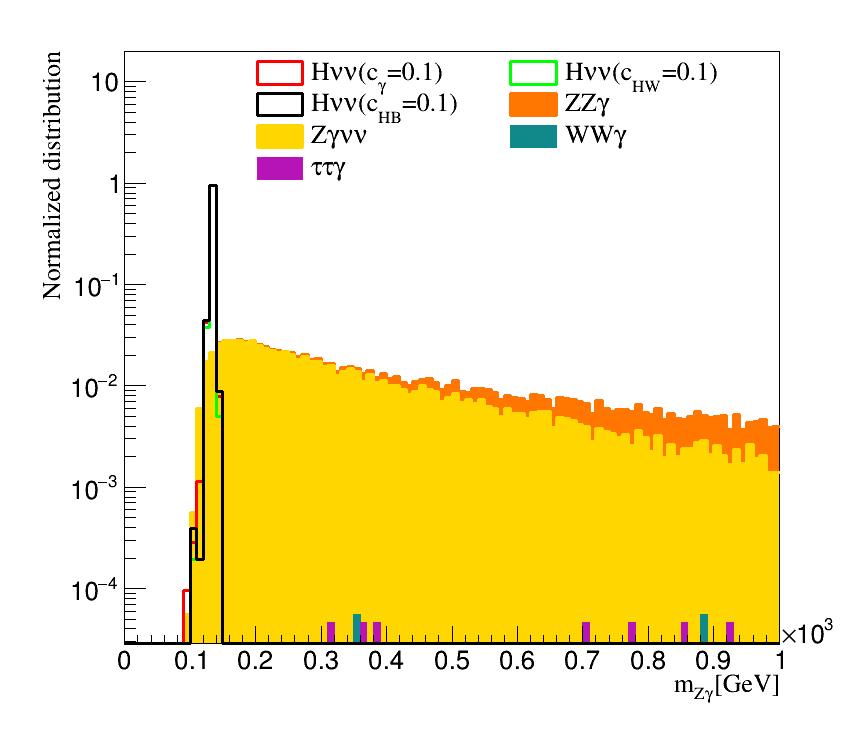}
	\caption{Normalized kinematic distributions of input observables utilized 
	for multivariate signal-background discrimination at $\sqrt{s} = 10\text{ TeV}$.}
	\label{fig:Variables-Distributions10}
\end{figure*}

To suppress SM background processes, we analyze key kinematic observables that exhibit distinct topological 
profiles between signal and background events. Based on these features, we pursue two complementary analysis 
strategies to extract bounds on the Wilson coefficients: a baseline cut-and-count approach anchored on the 
reconstructed Higgs mass window, and an optimized multivariate classification. The normalized distributions 
of these discriminating kinematic variables are illustrated in Figs.~\ref{fig:Variables-Distributions3} 
and~\ref{fig:Variables-Distributions10} for $\sqrt{s} = 3\text{ TeV}$ and $10\text{ TeV}$, respectively.

In the baseline cut-and-count strategy, the signal region is isolated within a narrow invariant mass window around the 
SM Higgs mass, $120\text{ GeV} < m_{Z\gamma} < 130\text{ GeV}$. The candidate Higgs invariant mass $m_{Z\gamma}$ is 
reconstructed by combining the primary $Z$-boson candidate ($m_{\ell\ell}$) with the highest-$p_T$ isolated photon in the event. 
Because the signal originates from a narrow physical resonance, signal events concentrate heavily within this window, 
broadened only by experimental detector resolution effects. Conversely, major non-resonant background continuum processes, 
such as irreducible $\nu\bar{\nu}Z\gamma$, exhibit smooth and broad $m_{Z\gamma}$ 
spectra extending into the high-mass tail. Imposing this selection significantly suppresses continuum backgrounds, 
maximizing the signal-to-background ratio ($S/B$).

\begin{table}[htbp]
\centering
\caption{Cross sections ($\sigma$ [fb]) at inclusive production level (prior to particle decays) and after sequential selection cuts for signal benchmarks ($c_\gamma, c_{HW}, c_{HB} = 0.1$) and individual SM backgrounds at $\sqrt{s} = 3\text{ TeV}$ and $10\text{ TeV}$. Here, Preselection cut includes object acceptance ($p_T \ge 20\text{ GeV}$, $|\eta| \le 2.4$) and lepton-photon angular separation ($\Delta R(\ell, \gamma) > 0.4$).}
\label{tab:cutflow_detailed}
\begin{tabular}{l cccc}
\hline\hline
Process & Production level  & Preselection  & $Z$-mass window  & Higgs-mass window  \\
\hline
\multicolumn{5}{c}{$\mathbf{\sqrt{s} = 3\text{ TeV}}$} \\
\hline
Signal ($c_{\gamma} = 0.1$)    & 176.50 & 2.655 & 2.623 & 2.496 \\
Signal ($c_{HW} = 0.1$) & 57.69  & 0.839 & 0.829 & 0.792 \\
Signal ($c_{HB} = 0.1$) & 34.69  & 0.518 & 0.511 & 0.481 \\
$Z\gamma\nu\bar{\nu}$    & 136.60 & 2.765 & 2.720 & 0.0867 \\
$WW\gamma$              & 27.84  & 0.159 & 0.0029 & $< 10^{-4}$ \\
$\tau\tau\gamma$        & 2.569  & 0.527 & 0.0182 & 0.0003 \\
$ZZ\gamma$              & 2.641  & 0.0546 & 0.0529 & 0.0001 \\
\hline
\multicolumn{5}{c}{$\mathbf{\sqrt{s} = 10\text{ TeV}}$} \\
\hline
Signal ($c_{\gamma} = 0.1$)    & 299.30 & 3.164 & 3.132 & 2.971 \\
Signal ($c_{HW} = 0.1$) & 97.85  & 1.027 & 1.016 & 0.972 \\
Signal ($c_{HB} = 0.1$) & 58.81  & 0.609 & 0.602 & 0.570 \\
$Z\gamma\nu\bar{\nu}$   & 276.70 & 4.345 & 4.273 & 0.1046 \\
$WW\gamma$              & 5.15   & 0.0098 & $1.03 \times 10^{-5}$ & 0 \\
$\tau\tau\gamma$        & 0.4296 & 0.0047 & $2.58 \times 10^{-6}$ & 0 \\
$ZZ\gamma$              & 0.4273 & 0.0029 & 0.0023 & $< 10^{-6}$ \\
\hline\hline
\end{tabular}
\end{table}

The cross sections at the inclusive production level (prior to particle decays) and following each sequential selection 
cut are summarized in Table~\ref{tab:cutflow_detailed} for $\sqrt{s} = 3\text{ TeV}$ and $\sqrt{s} = 10\text{ TeV}$. 
The final effective cross sections in the Higgs-mass window column serve as the direct input for extracting limits on 
the Wilson coefficients in the baseline cut-and-count analysis.

To exploit full multidimensional correlations and further enhance discrimination beyond single-variable cuts, 
we employ the Toolkit for Multivariate Data Analysis (TMVA) framework~\cite{TMVA:2007ngy,Speckmayer:2010zz} 
to train a Boosted Decision Tree (BDT) algorithm. All background channels are incorporated into the training stage, 
weighted according to their respective cross sections. To achieve optimal separation power while preventing overtraining, 
the BDT model is trained using the following kinematic input 
features (Figs~\ref{fig:Variables-Distributions3} and \ref{fig:Variables-Distributions10}):
\begin{itemize}
    \item $\Delta R(\ell^+, \ell^-)$: the angular separation between the two final-state charged leptons;
    \item $\Delta R(Z, \gamma)$: the angular separation between the reconstructed $Z$ boson and the candidate photon;
    \item $\cos\theta(Z, \gamma)$: the cosine of the opening angle between the $Z$ boson and the candidate photon;
    \item $E_T^{\text{miss}}$: the total missing transverse energy in the event;
    \item $m_{\ell\ell}$: the reconstructed invariant mass of the $Z$-boson candidate;
    \item $m_{Z\gamma}$: the reconstructed invariant mass of the Higgs-boson candidate ($Z\gamma$ system).
\end{itemize}

\begin{figure*}[htbp]
	\centering
	\includegraphics[width=0.45\textwidth]{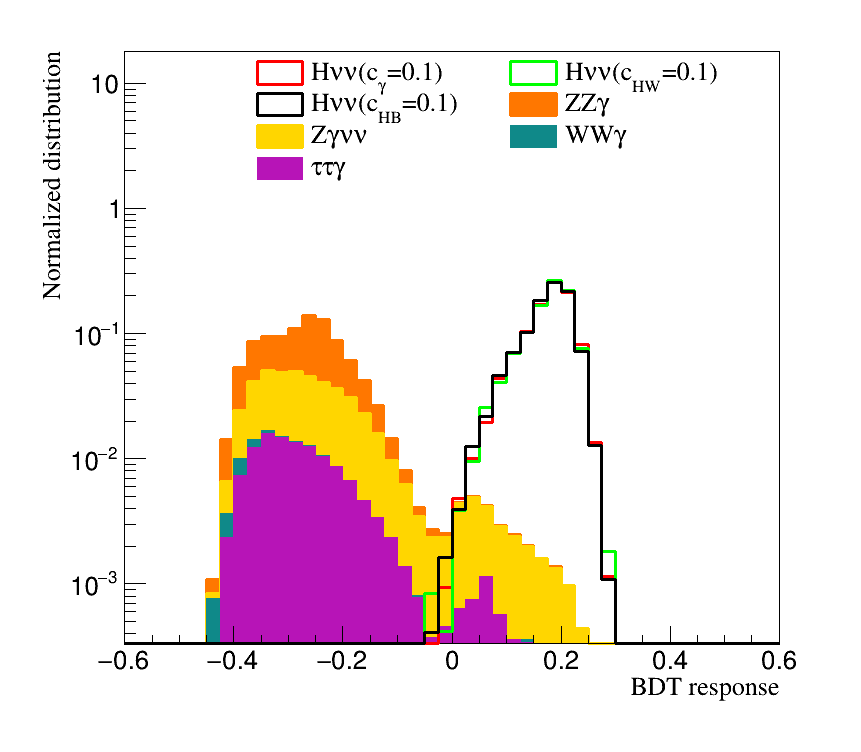}
	\includegraphics[width=0.45\textwidth]{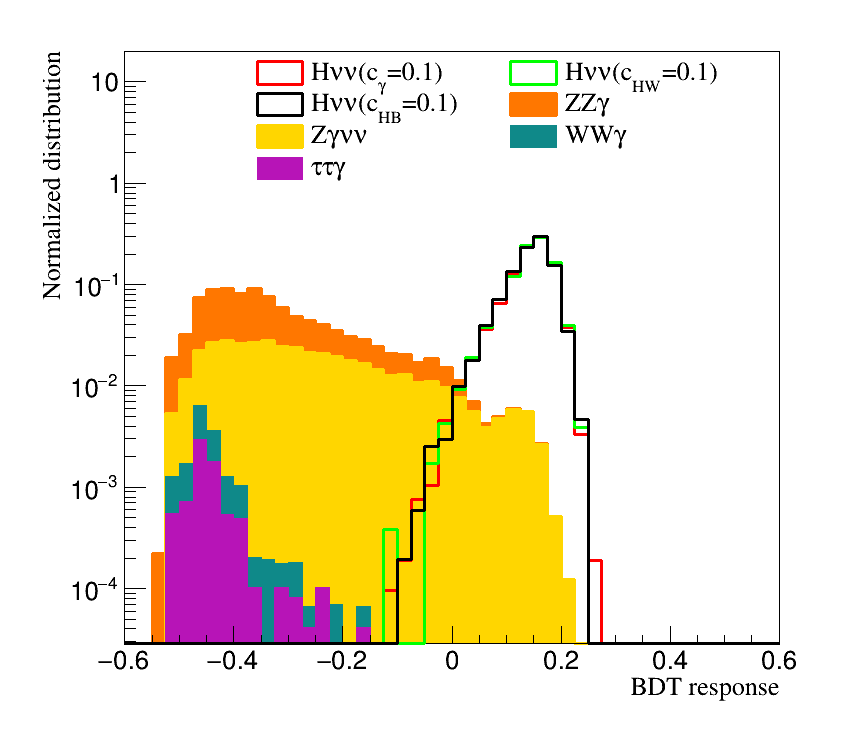}     
	\caption{BDT response distributions for the SMEFT signal benchmarks and total SM background 
	at $\sqrt{s} = 3\text{ TeV}$ (left) and $\sqrt{s} = 10\text{ TeV}$ (right).}
	\label{fig:BDT}
\end{figure*}

Applying the trained BDT models to the fully simulated event samples yields the response distributions shown 
in Fig.~\ref{fig:BDT} for $\sqrt{s} = 3\text{ TeV}$ and $10\text{ TeV}$. The BDT discriminant demonstrates strong 
separation capability, concentrating SM background events toward lower output scores while shifting signal distributions 
toward higher values. Compatibility between training and testing samples was verified via the Kolmogorov-Smirnov 
test to rule out overtraining. In the following section, these binned BDT distributions are utilized within 
a $\chi^2$ framework to derive 95\% CL bounds on the Wilson coefficients.

\section{Sensitivity estimation}\label{sec:Sensitivity}

In this section, we explore the sensitivity of a future high-energy muon collider to anomalous $hZ\gamma$ coupling
at $\sqrt{s} = 3\text{ TeV}$ ($1\text{ ab}^{-1}$) and $\sqrt{s} = 10\text{ TeV}$ ($10\text{ ab}^{-1}$). 
We do this by deriving expected 95\% confidence level (CL) upper limits on the relevant dimension-six Wilson 
coefficients: $c_{HW}$, $c_{HB}$, and $c_{\gamma}$. To get a comprehensive picture of the discovery potential, 
we break our analysis down into two complementary approaches. 
Subsection \ref{subsec:cut_and_count} walks through a straightforward cut-and-count strategy, 
where we extract our limits by focusing on the event yields that fall within a strict reconstructed Higgs mass window. 
Following that, Subsection \ref{subsec:bdt_analysis} takes a multivariate approach, leveraging a multi-binned shape 
analysis applied directly to the BDT output to better capture the kinematic differences between 
signal and background.

\subsection{Cut-and-Count Analysis}
\label{subsec:cut_and_count}

In the cut-and-count analysis, our primary strategy is to isolate the signal by applying a narrow invariant mass window 
cut around the SM Higgs boson, specifically requiring $120\text{ GeV} < m_{Z\gamma} < 130\text{ GeV}$. 
As demonstrated in the final column of Table~\ref{tab:cutflow_detailed}, this targeted cut efficiently suppresses 
the SM background while retaining a significant portion of the signal.

To translate these final event yields into physical limits, we utilize a standard $\chi^2$ statistical framework. 
The signal cross section's dependence on any given Wilson coefficient is parameterized quadratically as 
$S(c_i) = S_{\text{SM}} + c_i S_{\text{int}} + c_i^2 S_{\text{EFT}}$, where $S_{\text{SM}}$, $S_{\text{int}}$, 
and $S_{\text{EFT}}$ represent the SM expectation, the linear interference term, 
and the pure quadratic EFT contribution, respectively. Using the integrated event counts in our final signal 
region, we define our test statistic as~\cite{Barlow:1993dm,Cowan:2010js}:
\begin{equation}
\chi^2(c_i) = \frac{\left[S(c_i) - S_{\text{SM}}\right]^2}{B + (\delta_{\text{sys}} B)^2},
\end{equation}
where, $B$ stands for the total SM background yield, and $\delta_{\text{sys}}$ introduces a fractional 
systematic uncertainty, which we evaluate at realistic benchmarks of $5\%$ and $10\%$. Under the SM hypothesis, 
the expected one-dimensional bounds at the 95\% confidence level are determined by the interval 
where $\Delta\chi^2(c_i) = \chi^2(c_i) - \chi^2_{\text{min}} \le 3.841$, corresponding to one degree of freedom. 

Applying this procedure across both proposed energy stages, the resulting 95\% CL bounds on the Wilson 
coefficients at $\sqrt{s} = 3\text{ TeV}$ ($1\text{ ab}^{-1}$) and $\sqrt{s} = 10\text{ TeV}$ ($10\text{ ab}^{-1}$) 
are summarized in Table~\ref{tab:cut_and_count_limits}. As shown in the table, the upper limit on $c_{\gamma}$ is 
tighter than those of the other two couplings. This stronger bound is driven directly by the larger signal 
cross section associated with $c_{\gamma}$.
 
\begin{table}[htbp]
\centering
\caption{Expected 95\% CL limits on Wilson coefficients from the Cut-and-Count analysis 
at $\sqrt{s} = 3\text{ TeV}$ ($1\text{ ab}^{-1}$) and $\sqrt{s} = 10\text{ TeV}$ ($10\text{ ab}^{-1}$) 
under 5\% and 10\% systematic uncertainties.}
\label{tab:cut_and_count_limits}
\begin{tabular}{lcccc}
\hline\hline
 & \multicolumn{2}{c}{$\sqrt{s} = 3\text{ TeV}$ ($1\text{ ab}^{-1}$)} & 
 \multicolumn{2}{c}{$\sqrt{s} = 10\text{ TeV}$ ($10\text{ ab}^{-1}$)} \\
\cline{2-3} \cline{4-5}
Operator & 5\% Sys. & 10\% Sys. & 5\% Sys. & 10\% Sys. \\
\hline
$c_{HW}$ & $[-0.036, +0.009]$ & $[-0.037, +0.011]$ & $[-0.029, +0.005]$ & $[-0.032, +0.009]$ \\
$c_{HB}$ & $[-0.009, +0.035]$ & $[-0.011, +0.036]$ & $[-0.005, +0.029]$ & $[-0.009, +0.032]$ \\
$c_{\gamma}$    & $[-0.019, +0.005]$ & $[-0.020, +0.006]$ & $[-0.005, +0.003]$ & $[-0.019, +0.004]$ \\
\hline\hline
\end{tabular}
\end{table}

\subsection{BDT Multivariate Analysis}
\label{subsec:bdt_analysis}

To further enhance our sensitivity, we advance to a multivariate shape analysis. Instead of relying on 
a single mass window, we train a BDT to effectively separate the signal from the background and evaluate 
limits across its entire output distribution, as illustrated in Figure~\ref{fig:BDT}. To achieve this, 
we extend our previous statistical framework into a multi-binned $\chi^2$ analysis, summing over the 
80 output bins of the BDT discriminator:
\begin{equation}
\chi^2(c_i) = \sum_{k=1}^{80} \frac{\left[S_k(c_i) - S_{k,\text{SM}}\right]^2}{B_k + (\delta_{\text{sys}} B_k)^2 + \sigma_{\text{MC},k}^2},
\end{equation}
where $S_k(c_i)$ and $B_k$ denote the expected signal and background yields in the $k$-th bin. 
The term $\sigma_{\text{MC},k}^2 = \sum_{j} (L \cdot w_{j})^2$ accounts for the Monte Carlo statistical uncertainty, 
defined as the sum of squared per-event weights $w_j$ scaled to the target luminosity $L$. For bins containing zero 
simulated background events ($B_k = 0$), the expected background is substituted with the 68\% confidence level Poisson 
upper limit ($1.14$ events) scaled by the dominant background weight, ensuring a well-behaved denominator while 
avoiding unphysical zero-background singularities.
The parameter $\delta_{\text{sys}}$ introduces a fractional systematic uncertainty, which we evaluate 
at realistic benchmarks of $5\%$ and $10\%$. Under the SM hypothesis, 
the expected one-dimensional bounds at the 95\% confidence level are determined by the interval 
satisfying $\Delta\chi^2(c_i) = \chi^2(c_i) - \chi^2_{\text{min}} \le 3.841$, corresponding to one degree of freedom. 

The resulting 95\% CL limits on the Wilson coefficients extracted from this BDT multivariate analysis 
at $\sqrt{s} = 3\text{ TeV}$ ($1\text{ ab}^{-1}$) and $\sqrt{s} = 10\text{ TeV}$ ($10\text{ ab}^{-1}$) 
under both systematic uncertainty scenarios are summarized in Table~\ref{tab:bdt_limits}. Comparing the two methodologies, 
the multivariate BDT shape analysis improves the expected limits over the traditional cut-and-count approach by 
roughly $10\%$, demonstrating the distinct advantage of exploiting differential kinematic distributions rather 
than just integrated event counts. Furthermore, consistent with our earlier findings, the constraints 
on $c_\gamma$ remain slightly tighter than those on $c_{HW}$ and $c_{HB}$, driven by the larger 
signal cross section. Note that bounds at $\sqrt{s} = 3\text{ TeV}$ remain unchanged between $5\%$ and $10\%$ 
systematic uncertainties as the sensitivity in this regime is strictly statistics-limited.

\begin{table}[htbp]
\centering
\caption{Expected 95\% CL limits on Wilson coefficients from the BDT multivariate analysis at 
$\sqrt{s} = 3\text{ TeV}$ ($1\text{ ab}^{-1}$) and $\sqrt{s} = 10\text{ TeV}$ ($10\text{ ab}^{-1}$) 
under 5\% and 10\% systematic uncertainties.}
\label{tab:bdt_limits}
\begin{tabular}{lcccc}
\hline\hline
 & \multicolumn{2}{c}{$\sqrt{s} = 3\text{ TeV}$ ($1\text{ ab}^{-1}$)} & 
 \multicolumn{2}{c}{$\sqrt{s} = 10\text{ TeV}$ ($10\text{ ab}^{-1}$)} \\
\cline{2-3} \cline{4-5}
Operator & 5\% Sys. & 10\% Sys. & 5\% Sys. & 10\% Sys. \\
\hline
$c_{HW}$ & $[-0.033, +0.006]$ & $[-0.033, +0.006]$ & $[-0.005, +0.003]$ & $[-0.007, +0.004]$ \\
$c_{HB}$ & $[-0.006, +0.032]$ & $[-0.006, +0.032]$ & $[-0.003, +0.005]$ & $[-0.004, +0.007]$ \\
$c_{\gamma}$    & $[-0.018, +0.003]$ & $[-0.018, +0.003]$ & $[-0.002, +0.001]$ & $[-0.003, +0.002]$ \\
\hline\hline
\end{tabular}
\end{table}
	
These individual bounds on the Wilson coefficients can be translated directly into limits on the effective 
interaction coupling $g_{hZ\gamma}^{\text{eff}}$. Taking each operator one at a time, our $10~\mathrm{TeV}$ 
BDT analysis (with 5\% systematic uncertainty) yields the following projected $95\%$ CL intervals (using Eq.~\ref{eq:g_hzgamma}):
\begin{align}
g_{hZ\gamma}^{\text{eff}} &\in [-0.0016, \, +0.0027] \quad (\text{via } c_{HW}), \nonumber \\
g_{hZ\gamma}^{\text{eff}} &\in [-0.0016, \, +0.0027] \quad (\text{via } c_{HB}), \nonumber \\
g_{hZ\gamma}^{\text{eff}} &\in [-0.0036, \, +0.0018] \quad (\text{via } c_{\gamma}).
\end{align}
While $c_{HW}$ and $c_{HB}$ lead to a slightly narrower overall interval width, $c_\gamma$ sets the strongest 
limit on positive coupling deviations. Frame-independent quantities like $g_{hZ\gamma}^{\text{eff}}$ allow our 
results to be compared cleanly against LHC limits or future studies using different SMEFT bases.

A comparison with current bounds and projections across different collider eras highlights the exceptional reach of 
a high-energy muon collider. While LHC Run~2 data~\cite{ATLAS:2020zgamma,CMS:2022zgamma,ATLAS:2023yqk} currently bound 
$|g_{hZ\gamma}^{\text{eff}}|$ at the $\mathcal{O}(10^{-1})$ level and HL-LHC projections 
($14~\mathrm{TeV}, 3~\mathrm{ab}^{-1}$)~\cite{Cepeda:2019klc} expect to reach $\mathcal{O}(10^{-2})$, 
future $e^+e^-$ colliders such as FCC-ee~\cite{Abada:2019FCCee} and CLIC~\cite{deBlas:2018CLIC} 
project constraints in the $\mathcal{O}(10^{-2}\text{--}10^{-3})$ range. Our $10~\mathrm{TeV}$ 
muon collider BDT analysis tightens these bounds to $|g_{hZ\gamma}^{\text{eff}}| \lesssim 3 \times 10^{-3}$ at $95\%$ CL, 
surpassing HL-LHC sensitivity by nearly an order of magnitude and offering superior or highly competitive 
precision compared to FCC-ee and CLIC benchmarks.

\section{Summary}
\label{sec:summary}

In the Standard Model (SM), the decay of the Higgs boson into a $Z$ boson and a photon is a loop-induced process, 
resulting in a heavily suppressed branching fraction of approximately $(1.5 \pm 0.1) \times 10^{-3}$ for a 
$125\text{ GeV}$ Higgs mass. Because of this suppression, the $h \to Z\gamma$ channel serves as a highly 
sensitive probe for beyond the Standard Model (BSM) physics, as new heavy virtual particles running in 
the loop can significantly enhance the decay rate relative to the SM expectation. 

In this work, we investigated single Higgs boson production in association with neutrinos, followed by the 
$h \to Z\gamma$ decay, at a future high-energy muon collider. To parameterize potential new physics effects, 
we adopted the Standard Model Effective Field Theory (SMEFT) framework, focusing specifically on the 
dimension-six operators that modify the Higgs vertex: $\mathcal{O}_{HW}$, $\mathcal{O}_{HB}$, 
and $\mathcal{O}_{\gamma}$. 

We derived projected 95\% confidence level (CL) upper bounds on the corresponding Wilson coefficients 
($c_{HW}$, $c_{HB}$, and $c_{\gamma}$) using two complementary analytical strategies. The first method 
consisted of a traditional cut-and-count analysis, which isolated the signal by applying a targeted 
invariant mass cut around the reconstructed Higgs resonance. For the second approach, we employed the 
Toolkit for Multivariate Data Analysis (TMVA) framework to train a Boosted Decision Tree (BDT) algorithm. 
Both methodologies utilized a standard $\chi^2$ statistical framework to extract limits. However, while 
the cut-and-count method relied on a single-bin event yield, the BDT strategy leveraged a shape analysis 
across 80 bins of the discriminator output. By exploiting differential kinematic features, the multi-binned 
BDT approach successfully enhanced the sensitivity, yielding bounds that consistently outperformed the 
cut-and-count limits.

To highlight our most stringent projections, the multivariate BDT analysis at a $10\text{ TeV}$ 
muon collider (assuming an integrated luminosity of $10\text{ ab}^{-1}$ and a 5\% systematic uncertainty) 
yields expected 95\% CL intervals of $c_{HW} \in [-0.005, +0.003]$ and $c_{HB} \in [-0.003, +0.005]$. 
Notably, the bound on the $c_{\gamma}$ coefficient is slightly tighter due to its larger cross section 
enhancement, reaching $c_{\gamma} \in [-0.002, 0.001]$. Furthermore, our projected $95\%$ CL limits on 
$g_{hZ\gamma}^{\text{eff}}$ show that a $10~\mathrm{TeV}$ muon collider can restrict anomalous $hZ\gamma$ 
interactions down to the per-mille level ($\vert{}g_{hZ\gamma}^{\text{eff}}\vert{} \lesssim 3 \times 10^{-3}$), 
establishing a basis-independent benchmark for future precision Higgs studies.
	
\section*{Acknowledgments} \label{sec:ack}                                            
S.K. would like to thank the International Centre for Theoretical Physics (ICTP) for its kind hospitality 
during the final stages of this work. The authors acknowledge the use of Google's Gemini for assistance with language 
polishing and stylistic refinement. All scientific content, data analysis, and final conclusions remain entirely the 
responsibility of the authors.
	
\bibliographystyle{JHEP}                                                                                      
\bibliography{Draft}

@article{CMS:2012qbp,
    author = "Chatrchyan, Serguei and others",
    collaboration = "CMS",
    title = "{Observation of a New Boson at a Mass of 125 GeV with the CMS Experiment at the LHC}",
    eprint = "1207.7235",
    archivePrefix = "arXiv",
    primaryClass = "hep-ex",
    reportNumber = "CMS-HIG-12-028, CERN-PH-EP-2012-220",
    doi = "10.1016/j.physletb.2012.08.021",
    journal = "Phys. Lett. B",
    volume = "716",
    pages = "30--61",
    year = "2012"
}

@article{ATLAS:2012yve,
    author = "Aad, Georges and others",
    collaboration = "ATLAS",
    title = "{Observation of a new particle in the search for the Standard Model Higgs boson with the ATLAS detector at the LHC}",
    eprint = "1207.7214",
    archivePrefix = "arXiv",
    primaryClass = "hep-ex",
    reportNumber = "CERN-PH-EP-2012-218",
    doi = "10.1016/j.physletb.2012.08.020",
    journal = "Phys. Lett. B",
    volume = "716",
    pages = "1--29",
    year = "2012"
}

@article{ATLAS:2022vkf,
    author = "Aad, Georges and others",
    collaboration = "ATLAS",
    title = "{A detailed map of Higgs boson interactions by the ATLAS experiment ten years after the discovery}",
    eprint = "2207.00092",
    archivePrefix = "arXiv",
    primaryClass = "hep-ex",
    reportNumber = "CERN-EP-2022-057",
    doi = "10.1038/s41586-022-04893-w",
    journal = "Nature",
    volume = "607",
    number = "7917",
    pages = "52--59",
    year = "2022",
    note = "[Erratum: Nature 612, E24 (2022)]"
}

@article{CMS:2022dwd,
    author = "Tumasyan, Armen and others",
    collaboration = "CMS",
    title = "{A portrait of the Higgs boson by the CMS experiment ten years after the discovery.}",
    eprint = "2207.00043",
    archivePrefix = "arXiv",
    primaryClass = "hep-ex",
    reportNumber = "CMS-HIG-22-001, CERN-EP-2022-039",
    doi = "10.1038/s41586-022-04892-x",
    journal = "Nature",
    volume = "607",
    number = "7917",
    pages = "60--68",
    year = "2022",
    note = "[Erratum: Nature 623, (2023)]"
}

@article{Djouadi:1997yw,
    author = "Djouadi, A. and Kalinowski, J. and Spira, M.",
    title = "{HDECAY: A Program for Higgs boson decays in the standard model and its supersymmetric extension}",
    eprint = "hep-ph/9704448",
    archivePrefix = "arXiv",
    reportNumber = "DESY-97-079, IFT-96-29, PM-97-04",
    doi = "10.1016/S0010-4655(97)00123-9",
    journal = "Comput. Phys. Commun.",
    volume = "108",
    pages = "56--74",
    year = "1998"
}

@article{LHCHiggsCrossSectionWorkingGroup:2016ypw,
    author = "de Florian, D. and others",
    collaboration = "LHC Higgs Cross Section Working Group",
    title = "{Handbook of LHC Higgs Cross Sections: 4. Deciphering the Nature of the Higgs Sector}",
    eprint = "1610.07922",
    archivePrefix = "arXiv",
    primaryClass = "hep-ph",
    reportNumber = "CERN-2017-002-M, CERN-2017-002",
    doi = "10.23731/CYRM-2017-002",
    journal = "CERN Yellow Rep. Monogr.",
    volume = "2",
    pages = "1--869",
    year = "2017"
}

@article{Azatov:2013ura,
    author = "Azatov, Aleksandr and Contino, Roberto and Di Iura, Andrea and Galloway, Jamison",
    title = "{New Prospects for Higgs Compositeness in $h \to Z\gamma$}",
    eprint = "1308.2676",
    archivePrefix = "arXiv",
    primaryClass = "hep-ph",
    doi = "10.1103/PhysRevD.88.075019",
    journal = "Phys. Rev. D",
    volume = "88",
    number = "7",
    pages = "075019",
    year = "2013"
}

@article{Cao:2018cms,
    author = "Cao, Qing-Hong and Xu, Ling-Xiao and Yan, Bin and Zhu, Shou-Hua",
    title = "{Signature of pseudo Nambu{\textendash}Goldstone Higgs boson in its decay}",
    eprint = "1810.07661",
    archivePrefix = "arXiv",
    primaryClass = "hep-ph",
    doi = "10.1016/j.physletb.2018.12.040",
    journal = "Phys. Lett. B",
    volume = "789",
    pages = "233--237",
    year = "2019"
}

@article{Low:2011gn,
    author = "Low, Ian and Lykken, Joseph and Shaughnessy, Gabe",
    title = "{Singlet scalars as Higgs imposters at the Large Hadron Collider}",
    eprint = "1105.4587",
    archivePrefix = "arXiv",
    primaryClass = "hep-ph",
    reportNumber = "FERMILAB-PUB-11-312-T",
    doi = "10.1103/PhysRevD.84.035027",
    journal = "Phys. Rev. D",
    volume = "84",
    pages = "035027",
    year = "2011"
}

@article{Low:2012rj,
    author = "Low, Ian and Lykken, Joseph and Shaughnessy, Gabe",
    title = "{Have We Observed the Higgs (Imposter)?}",
    eprint = "1207.1093",
    archivePrefix = "arXiv",
    primaryClass = "hep-ph",
    reportNumber = "ANL-HEP-PR-12-49, FERMILAB-PUB-12-323-T, NUHEP-TH-12-09, ANL-HEP-PR-12-49; FERMILAB-PUB-12-323-T; NUHEP-TH/12-09",
    doi = "10.1103/PhysRevD.86.093012",
    journal = "Phys. Rev. D",
    volume = "86",
    pages = "093012",
    year = "2012"
}

@article{Carena:2012xa,
    author = "Carena, Marcela and Low, Ian and Wagner, Carlos E. M.",
    title = "{Implications of a Modified Higgs to Diphoton Decay Width}",
    eprint = "1206.1082",
    archivePrefix = "arXiv",
    primaryClass = "hep-ph",
    reportNumber = "ANL-HEP-PR-12-35, FERMILAB-PUB-12-922-T",
    doi = "10.1007/JHEP08(2012)060",
    journal = "JHEP",
    volume = "08",
    pages = "060",
    year = "2012"
}

@article{Chiang:2012qz,
    author = "Chiang, Cheng-Wei and Yagyu, Kei",
    title = "{Higgs boson decays to {\ensuremath{\gamma}}{\ensuremath{\gamma}} and Z{\ensuremath{\gamma}} in models with Higgs extensions}",
    eprint = "1207.1065",
    archivePrefix = "arXiv",
    primaryClass = "hep-ph",
    doi = "10.1103/PhysRevD.87.033003",
    journal = "Phys. Rev. D",
    volume = "87",
    number = "3",
    pages = "033003",
    year = "2013"
}

@article{Chen:2013vi,
    author = "Chen, Chian-Shu and Geng, Chao-Qiang and Huang, Da and Tsai, Lu-Hsing",
    title = "{New Scalar Contributions to $h\to Z\gamma$}",
    eprint = "1301.4694",
    archivePrefix = "arXiv",
    primaryClass = "hep-ph",
    doi = "10.1103/PhysRevD.87.075019",
    journal = "Phys. Rev. D",
    volume = "87",
    pages = "075019",
    year = "2013"
}

@article{ATLAS:2023yqk,
    author = "Aad, Georges and others",
    collaboration = "ATLAS, CMS",
    title = "{Evidence for the Higgs Boson Decay to a Z Boson and a Photon at the LHC}",
    eprint = "2309.03501",
    archivePrefix = "arXiv",
    primaryClass = "hep-ex",
    reportNumber = "CERN-EP-2023-157",
    doi = "10.1103/PhysRevLett.132.021803",
    journal = "Phys. Rev. Lett.",
    volume = "132",
    number = "2",
    pages = "021803",
    year = "2024"
}

@article{Cao:2015iua,
    author = "Cao, Qing-Hong and Wang, Hao-Ran and Zhang, Ya",
    title = "{Probing $HZ\gamma$ and $H\gamma\gamma$ anomalous couplings in the process $e^+e^- \to H\gamma$}",
    eprint = "1505.00654",
    archivePrefix = "arXiv",
    primaryClass = "hep-ph",
    doi = "10.1088/1674-1137/39/11/113102",
    journal = "Chin. Phys. C",
    volume = "39",
    number = "11",
    pages = "113102",
    year = "2015"
}

@article{No:2016ezr,
    author = "No, Jose Miguel and Spannowsky, Michael",
    title = "{A Boost to $h \to Z \gamma$: from LHC to Future $e^+ e^-$ Colliders}",
    eprint = "1612.06626",
    archivePrefix = "arXiv",
    primaryClass = "hep-ph",
    doi = "10.1103/PhysRevD.95.075027",
    journal = "Phys. Rev. D",
    volume = "95",
    number = "7",
    pages = "075027",
    year = "2017"
}

@article{Goertz:2019uek,
    author = "Goertz, Florian and Madge, Eric and Schwaller, Pedro and Tenorth, Valentin Titus",
    title = "{Discovering the $h\to Z \gamma$ decay in $t \bar t$ associated production}",
    eprint = "1909.07390",
    archivePrefix = "arXiv",
    primaryClass = "hep-ph",
    reportNumber = "MITP/19-061",
    doi = "10.1103/PhysRevD.102.053004",
    journal = "Phys. Rev. D",
    volume = "102",
    number = "5",
    pages = "053004",
    year = "2020"
}

@article{Durieux:2017rsg,
    author = "Durieux, Gauthier and Grojean, Christophe and Gu, Jiayin and Wang, Kechen",
    title = "{The leptonic future of the Higgs}",
    eprint = "1704.02333",
    archivePrefix = "arXiv",
    primaryClass = "hep-ph",
    reportNumber = "DESY-17-018",
    doi = "10.1007/JHEP09(2017)014",
    journal = "JHEP",
    volume = "09",
    pages = "014",
    year = "2017"
}

@article{Forslund:2023reu,
    author = "Forslund, Matthew and Meade, Patrick",
    title = "{Precision Higgs width and couplings with a high energy muon collider}",
    eprint = "2308.02633",
    archivePrefix = "arXiv",
    primaryClass = "hep-ph",
    doi = "10.1007/JHEP01(2024)182",
    journal = "JHEP",
    volume = "01",
    pages = "182",
    year = "2024"
}

@article{deLima:2024ybb,
    author = "de Lima, Carlos Henrique and Tuckler, Douglas",
    title = "{Sign of gauge-Higgs boson couplings at future lepton colliders}",
    eprint = "2408.14536",
    archivePrefix = "arXiv",
    primaryClass = "hep-ph",
    doi = "10.1103/PhysRevD.111.015021",
    journal = "Phys. Rev. D",
    volume = "111",
    number = "1",
    pages = "015021",
    year = "2025"
}

@article{Spor:2025ezn,
    author = "Spor, Serdar",
    title = "{Constraints on the anomalous Higgs boson couplings in Z{\ensuremath{\gamma}}{\ensuremath{\gamma}} production at muon collider}",
    eprint = "2501.11664",
    archivePrefix = "arXiv",
    primaryClass = "hep-ph",
    doi = "10.1016/j.nuclphysb.2025.117170",
    journal = "Nucl. Phys. B",
    volume = "1020",
    pages = "117170",
    year = "2025"
}

@article{Contino:2013kra,
    author = "Contino, Roberto and Ghezzi, Margherita and Grojean, Christophe and Muhlleitner, Margarete and Spira, Michael",
    title = "{Effective Lagrangian for a light Higgs-like scalar}",
    eprint = "1303.3876",
    archivePrefix = "arXiv",
    primaryClass = "hep-ph",
    reportNumber = "CERN-PH-TH-2013-047, KA-TP-06-2013, PSI-PR-13-04",
    doi = "10.1007/JHEP07(2013)035",
    journal = "JHEP",
    volume = "07",
    pages = "035",
    year = "2013"
}

@article{Alloul:2013naa,
    author = "Alloul, Adam and Fuks, Benjamin and Sanz, Ver{\'o}nica",
    title = "{Phenomenology of the Higgs Effective Lagrangian via FEYNRULES}",
    eprint = "1310.5150",
    archivePrefix = "arXiv",
    primaryClass = "hep-ph",
    reportNumber = "CERN-PH-TH-2013-248",
    doi = "10.1007/JHEP04(2014)110",
    journal = "JHEP",
    volume = "04",
    pages = "110",
    year = "2014"
}

@article{AlAli:2021let,
    author = "Al Ali, Hind and others",
    title = "{The muon Smasher{\textquoteright}s guide}",
    eprint = "2103.14043",
    archivePrefix = "arXiv",
    primaryClass = "hep-ph",
    doi = "10.1088/1361-6633/ac6678",
    journal = "Rept. Prog. Phys.",
    volume = "85",
    number = "8",
    pages = "084201",
    year = "2022"
}

@article{Aime:2022flm,
    author = "Aime, Chiara and others",
    title = "{Muon Collider Physics Summary}",
    eprint = "2203.07256",
    archivePrefix = "arXiv",
    primaryClass = "hep-ph",
    reportNumber = "FERMILAB-PUB-22-377-PPD",
    month = "3",
    year = "2022"
}

@article{Costantini:2020stv,
    author = "Costantini, Antonio and De Lillo, Federico and Maltoni, Fabio and Mantani, Luca and Mattelaer, Olivier and Ruiz, Richard and Zhao, Xiaoran",
    title = "{Vector boson fusion at multi-TeV muon colliders}",
    eprint = "2005.10289",
    archivePrefix = "arXiv",
    primaryClass = "hep-ph",
    reportNumber = "CP3-20-20, MCNET-20-12, VBSCAN-PUB-03-20",
    doi = "10.1007/JHEP09(2020)080",
    journal = "JHEP",
    volume = "09",
    pages = "080",
    year = "2020"
}

@article{Alwall:2011uj,
    author = "Alwall, Johan and Herquet, Michel and Maltoni, Fabio and Mattelaer, Olivier and Stelzer, Tim",
    title = "{MadGraph 5 : Going Beyond}",
    eprint = "1106.0522",
    archivePrefix = "arXiv",
    primaryClass = "hep-ph",
    reportNumber = "FERMILAB-PUB-11-448-T",
    doi = "10.1007/JHEP06(2011)128",
    journal = "JHEP",
    volume = "06",
    pages = "128",
    year = "2011"
}

@article{Bierlich:2022pfr,
    author = "Bierlich, Christian and others",
    title = "{A comprehensive guide to the physics and usage of PYTHIA 8.3}",
    eprint = "2203.11601",
    archivePrefix = "arXiv",
    primaryClass = "hep-ph",
    reportNumber = "LU-TP 22-16, MCNET-22-04, FERMILAB-PUB-22-227-SCD",
    doi = "10.21468/SciPostPhysCodeb.8",
    journal = "SciPost Phys. Codeb.",
    volume = "2022",
    pages = "8",
    year = "2022"
}

@article{deFavereau:2013fsa,
    author = "de Favereau, J. and Delaere, C. and Demin, P. and Giammanco, A. and Lema{\^\i}tre, V. and Mertens, A. and Selvaggi, M.",
    collaboration = "DELPHES 3",
    title = "{DELPHES 3, A modular framework for fast simulation of a generic collider experiment}",
    eprint = "1307.6346",
    archivePrefix = "arXiv",
    primaryClass = "hep-ex",
    doi = "10.1007/JHEP02(2014)057",
    journal = "JHEP",
    volume = "02",
    pages = "057",
    year = "2014"
}

@article{Degrande:2011ua,
    author = "Degrande, Celine and Duhr, Claude and Fuks, Benjamin and Grellscheid, David and Mattelaer, Olivier and Reiter, Thomas",
    title = "{UFO - The Universal FeynRules Output}",
    eprint = "1108.2040",
    archivePrefix = "arXiv",
    primaryClass = "hep-ph",
    reportNumber = "CP3-11-25, IPHC-PHENO-11-04, IPPP-11-39, DCPT-11-78, MPP-2011-68",
    doi = "10.1016/j.cpc.2012.01.022",
    journal = "Comput. Phys. Commun.",
    volume = "183",
    pages = "1201--1214",
    year = "2012"
}

@article{Accettura:2023,
    author = "Accettura, C. and others",
    collaboration = "International Muon Collider",
    title = "{Towards a Muon Collider}",
    eprint = "2303.08533",
    archivePrefix = "arXiv",
    primaryClass = "hep-ex",
    doi = "10.1140/epjc/s10052-023-11889-x",
    journal = "Eur. Phys. J. C",
    volume = "83",
    number = "9",
    pages = "864",
    year = "2023"
}

@misc{MuonColliderDelphesCard,
    author = "{International Muon Collider Collaboration}",
    title = "{Delphes card for future Muon Collider}",
    howpublished = "\url{https://github.com/delphes/delphes/tree/master/cards/MuonCollider}",
    year = "2023"
}

@article{TMVA:2007ngy,
    author = "Hocker, Andreas and others",
    collaboration = "TMVA",
    title = "{TMVA - Toolkit for Multivariate Data Analysis}",
    eprint = "physics/0703039",
    archivePrefix = "arXiv",
    reportNumber = "CERN-OPEN-2007-007",
    month = "3",
    year = "2007"
}

@article{Speckmayer:2010zz,
    author = "Speckmayer, P. and Hocker, A. and Stelzer, J. and Voss, H.",
    editor = "Gruntorad, Jan and Lokajicek, Milos",
    title = "{The toolkit for multivariate data analysis, TMVA 4}",
    doi = "10.1088/1742-6596/219/3/032057",
    journal = "J. Phys. Conf. Ser.",
    volume = "219",
    pages = "032057",
    year = "2010"
}

@article{Barlow:1993dm,
    author = "Barlow, Roger J. and Beeston, Christine",
    title = "{Fitting using finite Monte Carlo samples}",
    reportNumber = "MAN-HEP-93-1",
    doi = "10.1016/0010-4655(93)90005-W",
    journal = "Comput. Phys. Commun.",
    volume = "77",
    pages = "219--228",
    year = "1993"
}

@article{Cowan:2010js,
    author = "Cowan, Glen and Cranmer, Kyle and Gross, Eilam and Vitells, Ofer",
    title = "{Asymptotic formulae for likelihood-based tests of new physics}",
    eprint = "1007.1727",
    archivePrefix = "arXiv",
    primaryClass = "physics.data-an",
    doi = "10.1140/epjc/s10052-011-1554-0",
    journal = "Eur. Phys. J. C",
    volume = "71",
    pages = "1554",
    year = "2011",
    note = "[Erratum: Eur.Phys.J.C 73, 2501 (2013)]"
}

@article{ATLAS:2020zgamma,
    author = "Aad, Georges and others",
    collaboration = "ATLAS",
    title = "{A search for the $Z\gamma$ decay mode of the Higgs boson in $pp$ collisions at $\sqrt{s} = 13$ TeV with the ATLAS detector}",
    eprint = "2005.05382",
    archivePrefix = "arXiv",
    primaryClass = "hep-ex",
    reportNumber = "CERN-EP-2020-061",
    doi = "10.1016/j.physletb.2020.135754",
    journal = "Phys. Lett. B",
    volume = "809",
    pages = "135754",
    year = "2020"
}

@article{CMS:2022zgamma,
    author = "Tumasyan, Armen and others",
    collaboration = "CMS",
    title = "{Search for Higgs boson decays to a Z boson and a photon in proton-proton collisions at $\sqrt{s} = 13$ TeV}",
    eprint = "2204.12945",
    archivePrefix = "arXiv",
    primaryClass = "hep-ex",
    reportNumber = "CMS-HIG-19-014, CERN-EP-2022-040",
    doi = "10.1007/JHEP05(2023)233",
    journal = "JHEP",
    volume = "05",
    pages = "233",
    year = "2023"
}

@article{Cepeda:2019klc,
    author = "Cepeda, M. and others",
    editor = "Dainese, Andrea and Mangano, Michelangelo and Meyer, Andreas B. and Nisati, Aleandro and Salam, Gavin P. and Vesterinen, Mika A.",
    title = "{Report from Working Group 2}: {Higgs Physics at the HL-LHC and HE-LHC}",
    eprint = "1902.00134",
    archivePrefix = "arXiv",
    primaryClass = "hep-ph",
    reportNumber = "CERN-LPCC-2018-04",
    doi = "10.23731/CYRM-2019-007.221",
    journal = "CERN Yellow Rep. Monogr.",
    volume = "7",
    pages = "221--584",
    year = "2019"
}

@article{Abada:2019FCCee,
    author = "Abada, A. and others",
    collaboration = "FCC",
    title = "{FCC-ee: The Lepton Collider}: {Future Circular Collider Conceptual Design Report Volume 2}",
    doi = "10.1140/epjc/s10052-019-6904-3",
    journal = "Eur. Phys. J. C",
    volume = "79",
    number = "6",
    pages = "474",
    year = "2019"
}

@article{deBlas:2018CLIC,
    author = "de Blas, J. and Durieux, G. and Grojean, C. and Guintrand, J. and Paul, A.",
    title = "{On the physics potential of $e^+e^-$ linear colliders}",
    eprint = "1907.04311",
    archivePrefix = "arXiv",
    primaryClass = "hep-ph",
    reportNumber = "DESY-19-118, HU-EP-19/20",
    doi = "10.1007/JHEP12(2019)117",
    journal = "JHEP",
    volume = "12",
    pages = "117",
    year = "2019"
}

\end{document}